\documentclass[manuscript]{acmart}

\usepackage{tabularx}
\usepackage{hyperref}
\usepackage[capitalise]{cleveref}
\usepackage{float}

\AtBeginDocument{%
  }

\setcopyright{none}
\begin{document}
\raggedbottom

\title{A Paradigm for Creative Ownership}

\author{Tejaswi Polimetla} 
\affiliation{%
  \institution{Harvard University}
  \city{Cambridge}
  \state{Massachusetts} 
  \country{USA}}
\email{tpolimetla@seas.harvard.edu}

\author{Katy Ilonka Gero}
\affiliation{%
  \institution{University of Sydney}
  \city{Sydney}
  \country{Australia}}
\email{katy.gero@sydney.edu.au}

\author{Elena L. Glassman} 
\affiliation{%
  \institution{Harvard University}
  \city{Cambridge}
  \state{Massachusetts} 
  \country{USA}}
\email{glassman@seas.harvard.edu}

\begin{abstract}
As generative AI tools become embedded in creative practice, questions of ownership in co-creative contexts are pressing. Yet studies of human-AI collaboration often invoke "ownership" without definition: sometimes conflating it with other concepts, and other times leaving interpretation to participants. This inconsistency makes findings difficult to compare across or even within studies. We introduce a framework of creative ownership comprising three dimensions - Person, Process, and System - each with three subdimensions, offering a shared language for both system design and HCI research. In semi-structured interviews with 21 creative professionals, we found that participants’ initial references to ownership (e.g., embodiment, control, concept) were fully encompassed by the framework, demonstrating its coverage. Once introduced, however, they also articulated and prioritized the remaining subdimensions, underscoring how the framework expands reflection and enables richer insights. Our contributions include 1) the framework, 2) a web-based visualization tool, and 3) empirical findings on its utility.
\end{abstract}



\maketitle

\section{Introduction}

Generative AI tools are now routine parts of a creative practice for many people—from drafting text and composing images to arranging sound and editing video. As teams, platforms, and models co-produce artifacts, creators regularly ask: \emph{what does it mean for this to feel "mine"?}  While questions of ownership have been asked since antiquity, and collaborations of all kinds can challenge our notions of ownership, generative AI requires us to revisit these questions because these new kinds of "collaboration" do not map cleanly onto either a human partner nor a simple tool. While law offers formal answers about rights and attribution, legal rights and psychological feelings do not go hand in hand, and Human-Computer Interaction research must grapple with \emph{creative ownership} as a lived, situated phenomenon. 

In their 2003 paper, \citet{Pierce2003} define psychological ownership as "that state where an individual feels as though the target of ownership or a piece of that target is ‘theirs’." In this paper, we will focus on a narrower definition revolving around \textit{creative ownership} in which the target of ownership is a creative product or artifact that the individual in question had a role in creating --- no matter how small or large. While prior HCI work engages in challenges involving creative ownership, the concept is not defined consistently. Some studies conflate ownership with adjacent ideas (e.g., control, vision, identity); others elicit participants’ views without a common scaffold, making results hard to compare across settings and media. Existing instruments either center creativity without ownership (e.g., Creativity Support Index \cite{cherry2014quantifying}) or synthesize ownership broadly but without a lens for creative work \cite{kuzminykh2020bemine}. As a result, researchers and designers lack a compact, comparable way to \emph{measure}, \emph{reason about}, and \emph{design for} creative ownership in human–AI collaboration.

Building upon literature across psychology, philosophy, the humanities and social sciences more broadly, and within human-computer interaction, we introduce a nine-subdimension framework of creative ownership organized across \emph{Person}, \emph{Process}, and \emph{System}. \emph{Person} captures how the artifact relates to the self; \emph{Process} characterizes the decisions, intentionality, and effort by which it is created; \emph{System} situates creation within its material, collaborative, and contextual conditions. We operationalize the framework as an interactive web tool to make the dimensions legible to users, and to standardize comparison across projects and fields. 

We conducted semi-structured interviews with 21 creative professionals across a diverse range of fields. We used a two-phase, within-participant protocol. Participants first described one high-ownership and one low-ownership project without the framework, then used our instrument to rate both works and reflect on the output. Qualitatively, pre-framework talk concentrated on a limited subset of subdimensions (embodiment, control, abstraction). Once introduced, participants articulated and prioritized \emph{all nine} subdimensions, enabling finer distinctions (e.g., conceptual authorship vs. physical production) and revealing medium-dependent nuances. Quantitatively, all nine subdimensions were consistently higher for high-ownership projects than for low-ownership ones; low-ownership ratings showed greater variance, indicating multiple pathways by which ownership can be diminished. Participants reported that the framework was a useful tool for reflection.

\noindent
This paper makes the following contributions:
\begin{itemize}
  \item \textbf{A concise framework} for creative ownership with nine subdimensions that span Person, Process, and System, grounded in psychology, philosophy, and HCI literature.
  \item \textbf{An interactive webtool} that enables elicitation, comparison, and visualization of ownership across projects and creative fields.
  \item \textbf{Empirical validation} via interviews with 21 creative professionals: (i) quantitative comparison across high and low ownership conditions and (ii) an understanding of how ownership is conceptualized both with and without the presence of the framework.
\end{itemize}

\section{Ownership in the HCI Literature}

While considerable work has been done at the intersection of AI and ownership in the HCI literature, there remains little consensus on what ownership entails or how it should be addressed across studies. Researchers often either leave it to participants to define ownership in their own terms, or conflate ownership with adjacent concepts or with a single strand of ownership.

Examples illustrate the breadth of interpretations. One study of artists’ perspectives on AI-generated art focuses primarily on the legal dimensions of ownership, raising questions of disclosure, threats to human labor, and intellectual property \cite{lovato2024foregrounding}. In workplace communication, \citet{kadoma2024role} conceptualize ownership as "mineness". \citet{draxler2024aighostwriter} and \citet{liang2025synthia} adopt a similar approach, asking questions such as "to whom should this text belong" or "I feel like this writing was purely produced by me" in the context of AI-supported writing. Others examine content ownership specifically in the setting of LLM-based writing assistants \cite{wasi2024llms, he2024ai}. Some researchers employ more structured methods: \citet{joshi2025writing}, for instance, ground their work in psychological ownership, asking about personal ownership, responsibility, personal connection, and emotional attachment. While this captures psychological dimensions, it does not address creativity. Conversely, \citet{wu2023owndiffusion} adapt the Creativity Support Index, which highlights creative aspects but not ownership.

Other work conflates ownership with related constructs, or dimensions that might be considered only part of ownership. \citet{weber2025drawing}, for example, use the term "artistic ownership" in studying support for creative goals, yet operationalize it through adjacent concepts such as creative vision, intentions, collaboration, pride, control, and emotional response \cite{weber2025drawing}. Even when researchers begin with a focused definition, as in Wasi et al.'s work~\cite{wasi2024llms} on content ownership, related ideas often surface—embodiment, identity, originality, and effort among them.

Still others leave ownership undefined, inviting participants to articulate their own interpretations. In study of AI-driven scriptwriting by \citet{weber2024wraiter}, participants associated ownership with ease, expression, collaboration, uniqueness, and enjoyment. Guo et al.’s work~\cite{guo2024exploring} on AI value alignment in collaborative ideation found participants emphasizing concept, originality, and creative agency. Similarly, Biermann et al.’s study~\cite{biermann2022fromtool} of AI-supported story writing reported themes such as plot, style, control, concept, and collaboration.

Efforts have been made in HCI to establish more unified frameworks, though these remain limited in scope. Two key examples are Cherry et al.’s Creativity Support Index and Kuzminykh et al.’s synthesis of ownership research in HCI \cite{cherry2014quantifying, kuzminykh2020bemine}. \citet{cherry2014quantifying} identify six dimensions of creativity support, yet the framework does not establish a direct connection to ownership. Conversely, \citet{kuzminykh2020bemine} provide a broad overview of ownership literature, but their synthesis is not readily suited to creative contexts. Building on these efforts, our aim is to develop a framework for ownership that is specifically tailored to creative practice and designed for use in HCI research.

\section{Literature Review}

The concept of ownership has been studied widely across the humanities and social sciences, with each field having a unique approach to the subject matter \cite{Pipes2000}. This literature review focuses on the fields of philosophy and psychology as these two fields most readily discuss issues related to how humans develop feelings of ownership on an individual level. Other social sciences such as economics, politics, and sociology are concerned with broader system level concerns. While other humanistic fields such as art and literature do discuss individual level concerns, the arguments developed are often grounded in formal philosophical reasoning. 

\subsection{Philosophy}

While there are many schools of philosophical thought that could be used to frame a discussion of ownership, two juxtaposing ones that encompass the duality of ownership related values are materialism and idealism. Materialist theories stem from notions of property as control over material entities, going as far as to stipulate that physical, material states are the ultimate determinants of reality, taking precedence over thought, consciousness, and abstract entities \cite{Melnyk2012, Smart1963}. On the contrary, idealism posits that something mental is the ultimate foundation of reality, and idealist theories of property and personhood are concerned with symbolic and mental conceptions of ownership \cite{GuyerHorstmann2015}. This dualistic framing captures both the tangible and intangible elements of ownership. 

One of the most fundamental materialist theories is Locke’s labor theory, which posits that "every man has a property in his own person," and thereby goes on to argue that when one mixes their labor with natural resources, the resulting good becomes their property - evoking the embodiment theory of personhood \cite{Locke1689, Radin1982}. "Bundle of Rights" views hold ownership as a set of contractual obligations between people in relation to property \cite{Waldron1985, SEPProperty, Honore1961}. More broadly, these theories lend themselves as a basis for the construction of property systems as they tend to be grounded in tangible, enforceable, socially structured claims. In contrast, idealist theories tend to be much more abstract. For instance, Hegel’s ideas of ownership stem from the notion that the "will" can be embodied in external entities, and that this embodiment is necessary for one’s actualization as a person cannot come to exist without both relation to and differentiation from the external environment \cite{Radin1982}. While the specifics of theories vary, the investment of one’s self, values, and identity as a means of developing feelings of ownership is a common theme that arises \cite{Sartre1943, Durkheim1957}.

Although these views are contrasting, both reveal key features of ownership such as the investment of labor and the reflection of identity that contribute to a broader conception and intuition of how feelings of ownership are developed. It is evident that both the material and immaterial have appealed to human sensibilities throughout history. 

\subsection{Psychology}

In the field of psychology, there have been numerous theoretical propositions and empirical studies attempting to explain the formation of psychological ownership. Several scholars have created frameworks based on decades of psychological research that capture key themes that have emerged time and again such as effectance and control of possessions \cite{Furby1978, White1959, McClelland1951}, positive affect \cite{Furby1978}, and symbolic meaning and personhood \cite{RochbergHalton1980}. These frameworks span a range of formulations ranging from Targets-Antecedents-Consequences-Interventions \cite{Lyu2023} to corrective dual-process models \cite{Morewedge2021}, among others \cite{Boyer2023, Pierce2001}. Some of the major themes found across frameworks include responsibility, accountability, identity, self-efficacy, belongingness, control, self-congruity, psychological closeness, object-knowledge, self-investment, and rights over the object. While these frameworks capture numerous dimensions of psychological ownership, they fall short in creative contexts, where ownership by creation begins to deviate from other forms of ownership. Research on the self-creation effect illustrates how creating something oneself can lead to stronger object valuation and a more profound sense of ownership - aspects that are often overlooked by traditional frameworks of ownership \cite{koster2015beliefs, brunneder2018self}. Therefore, we draw upon existing frameworks and approaches to produce a framework that is more streamlined for creative contexts.

In addition to these framework based approaches, there has also been much empirical work to understand the emergence of psychological ownership under particular circumstances. In a study of psychological ownership in creative work, Rouse identifies differences in task type (managerial, conceptual, production), task target (abstract, concrete), and task coordination (independent, autonomous) \cite{Rouse2013}. In a study of ePortfolio creation in online learning environments, researchers found that psychological ownership is correlated with perceived control of content, planning, personal data and access rights \cite{Buchem2012}. \citet{Levene2015} examined the effects of creation and intent on ownership judged and found that the effects of creation hold even when controlling for other factors. They also showed that successful and intentional creations are ascribed more ownership than unsuccessful or unintentional creations, and that creation is ascribed more ownership than the equivalent labor. A study of locus of control and psychological ownership found internal locus of control to be positively related to both effectance and self-identity motives \cite{McIntyre2009}. When measuring ownership sentiments in academic writing contexts, \citet{Nicholes2015} found that ownership sentiments tend to be significantly higher in creative genres as opposed to academic genres, and higher for argumentative papers as compared to research papers. Struggle is another theme that appears, with productive and creative struggle arising as both educational and design goals \cite{Fox2023, Zhou2023}.

\section{Framework}
\noindent Our framework consists of nine dimensions of creative ownership:
\begin{itemize}
    \item \textbf{\textcolor{red!70!black}{Person}}:
        \begin{itemize}
            \item \textbf{Embodiment} – How much do you feel that the finished product embodies your values, personality, and identity?
            \item \textbf{Occupancy} – How often do you feel preoccupied with thinking about the product?
            \item \textbf{Recognition} – Do you receive a lot of external recognition for your work?
        \end{itemize}
    \item \textbf{\textcolor{green!50!black}{Process}}:
        \begin{itemize}
            \item \textbf{Control} – How much control did you have over the creative decisions that shaped the final product?
            \item \textbf{Intentionality} – How intentional were you about the creative decisions that you made?
            \item \textbf{Effort} – How much physical and/or mental effort did you put into this work?
        \end{itemize}
    \item \textbf{\textcolor{blue!70!black}{System}}:
        \begin{itemize}
            \item \textbf{Production} – How closely did you physically work with the final product?
            \item \textbf{Abstraction} – How closely did you conceptually work with the final product?
            \item \textbf{Interdependence} – How independent were you during the creative process?
        \end{itemize}
\end{itemize}
The proposed framework incorporates the material and immaterial components of ownership, tying together elements of individual values and personhood, the creative process, and the systems in which these individuals and processes are embedded. \textbf{Personhood} in a creative context is conceptualized as how the self relates to the creative outcome. How does the creative product embody one’s identity, self and values, how consistently does it occupy the mind of the creator, and to what extent do others recognize the creator (and their personhood) for their work? The creative \textbf{process} is the series of steps by which the creator’s personhood is realized in the creation, and is characterized by the intentionality of the creator, the amount of control they have over the process, and the amount of effort - both mental and physical - that goes into the realization. The \textbf{system} refers to the environment in which this creative process is enacted and its dimensions are borrowed from Rouse’s findings \cite{Rouse2013}. Abstraction refers to the creator's role in conceptualizing the work, while production relates to their responsibility for its materialization. Interdependence reflects the creator's reliance on others or tools to complete the creation. While process and system dimensions may be correlated at times, this is not always the case as shown in \Cref{sec:process-vs-system}, which illustrates how these dimensions can vary through a series of case studies.

\begin{figure}[h]
    \centering 
    \includegraphics[width=\textwidth]{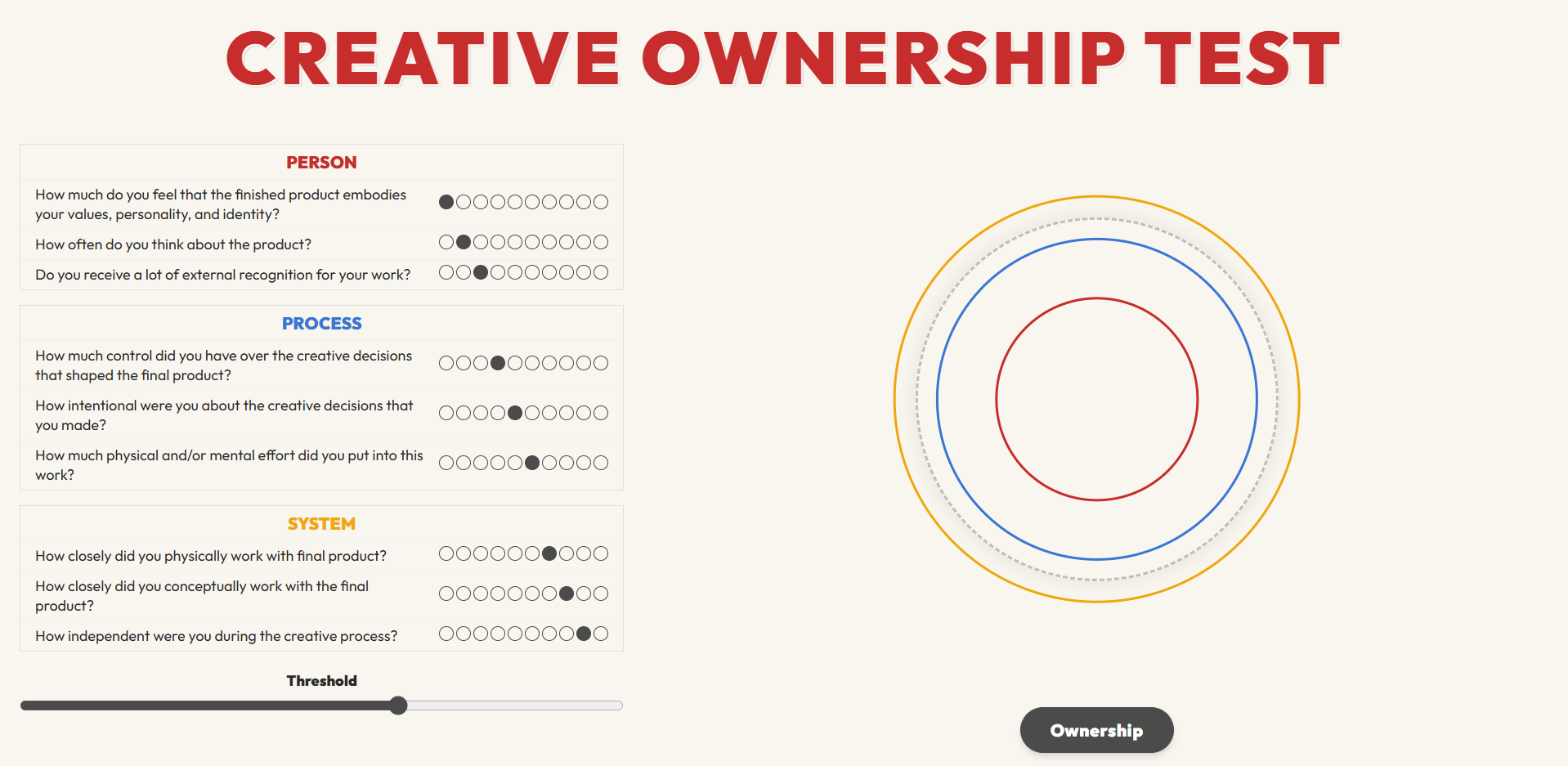} 
    \caption{Creative Ownership Test Webtool}
    \label{fig:framework}
\end{figure}

We choose to visualize this framework using three concentric circles, each representing one of the elements listed above, with each element composed of sub-dimensions. An increase in one dimension leads not only to an increase in the size of its corresponding circle, but also in the total size of the three circles together. Importantly, the visualization refrains from centering on any one element. Instead, precedence is context dependent, and the innermost circle can be thought of as the limiting factor in developing feelings of ownership. A threshold slider illustrates how individuals may have different thresholds for developing such feelings. The visualization can be explored using this interactive webtool (\Cref{fig:framework}): \href{https://creative-ownership-test.vercel.app/}{\textbf{Creative Ownership Test}}.
A simple example of the framework's application is the comparison between a master painter, like Picasso or Van Gogh, and a hobbyist art student (visualized in \Cref{sec:applications}). The master painter is widely recognized, with their work embodying their personhood. They invested significant time, intentionality, control, and effort into their art, working directly with the final product and maintaining independence throughout the creative process. In contrast, the hobbyist art student, following a paint-along workshop, is unlikely to view the finished painting as embodying their values. They have little control or intentionality in the outcome, exert moderate effort, and are highly reliant on the instructor. While they closely work with the final product, they play a minimal role in its conceptualization. From the figure, we can see that their main ownership "axis" is that of being part of a system of creation—they enact a small but critical part of the painting creation. Refer to \Cref{sec:examples} for more examples. 

Additionally, we map the primary concepts discussed in the existing literature to the concepts proposed by our framework. It appears that within this sample of ten papers, Embodiment, Abstraction, Interdependence, and Control appear most often, with some mention of Effort and Intentionality, and no mention of Occupancy, Recognition, or Production. This showcases how this new framework covers the existing concepts that are frequently elicited in the HCI literature, while also proposing new dimensions based on findings from psychology and philosophy that could contribute to a more comprehensive understanding of ownership. 

\begin{table}[h]
\centering
\caption{Mapping of existing ownership concepts to framework concepts}
\label{tab:papers-concepts}
\begin{tabularx}{\linewidth}{|l|X|X|}
    \hline
    \textbf{Papers} & \textbf{Paper Concepts} & \textbf{Framework Concepts} \\
    \hline
    \citet{kadoma2024role} & "Mineness" & Embodiment, Interdependence \\
    \hline
    \citet{draxler2024aighostwriter} & "To whom should the text belong" & Interdependence\\
    \hline
    \citet{liang2025synthia} & "The writing was purely produced by me" & Interdependence\\
    \hline
    \citet{he2024ai} & Content ownership & Abstraction \\
    \hline
    \citet{wasi2024llms} & Content ownership, originality & Abstraction \\
    & Embodiment, identity & Embodiment \\
    & Effort & Effort \\
    \hline
    \citet{joshi2025writing} & Personal ownership, personal connection, emotional attachment & Embodiment \\
    \hline
    \citet{weber2025drawing} & Creative vision & Abstraction \\
    & Intentions & Intentionality \\
    & Collaboration & Interdependence \\
    & Control & Control \\
    & Emotional response & Embodiment \\
    \hline
    \citet{weber2024wraiter} & Ease & Effort \\
    & Expression, uniqueness & Abstraction \\
    & Collaboration & Interdependence \\
    \hline
    \citet{guo2024exploring} & Concept, originality & Abstraction \\
    & Creative agency & Control \\
    \hline
    \citet{biermann2022fromtool} & Plot, concept & Abstraction \\
    & Style & Embodiment \\
    & Control & Control \\
    & Collaboration & Interdependence \\
    \hline
\end{tabularx}
\end{table}

\section{Validation Study --- Methodology}

Our methodological design was guided by the goal of comparing how participants described ownership before and after being introduced to the framework, with a focus on understanding the coverage and utility of the framework's dimensions. To capture this contrast, we asked them to reflect on both a high-ownership and a low-ownership creative project, enabling comparison across contexts as well as within individual experience. We refer to these phases as the pre-webtool and post-webtool sections of the study.

\subsection{Participants}

Potential participants were identified through a combination of referrals from the researchers’ professional networks, publicly available sources, and local art communities in the Greater Boston area. To be eligible, participants were required to: (1) work or participate significantly in a creative field, (2) have at least two finished creative products---one associated with high feelings of ownership and one with low feelings of ownership, (3) be fluent in English, and (4) be over 18 years of age. We recruited 20 participants via word of mouth, email, and snowball sampling.

Participants represented a wide range of creative domains (enumerated in \Cref{tab:participants}), and varied in experience from a few years of part-time practice to decades of dedicated mastery. This diversity was intentional, reflecting our goal of examining creative ownership across domains rather than within a single practice.

The study protocol was approved by our institutional ethics review board (IRB). All participants provided informed consent prior to participation. Each received \$25 in compensation, either as cash or a gift card. To protect anonymity while maintaining contextual richness, we refer to participants using pseudonyms with a field descriptor (e.g., P1, ukulelist, singer).
\begin{table*}[h]
\centering
\caption{Study Participants and Project Descriptions}
\label{tab:participants}
\begin{tabularx}{\linewidth}{|c|l|X|X|}
    \hline
    \textbf{ID} & \textbf{Field} & \textbf{High Ownership Project} & \textbf{Low Ownership Project} \\
    \hline
    P1 & Social Media & Content for personal Instagram account & Social media ad campaign \\
    \hline
    P2 & Ukulelist, Singer & Composed original song & Recorded some ukulele for another song's production \\
    \hline
    P3 & Dancer & Independent dance project for thesis & Choreography for a high school play \\
    \hline
    P4 & Nonfiction Writer & Collection of personal essays for thesis & Experimental art project in new visual medium \\
    \hline
    P5 & Architect & City planning project undertaken in academic setting & Professional architecture project \\
    \hline
    P6 & Video Game Developer & Master's degree capstone project video game & Game system designer at a university \\
    \hline
    P7 & Novelist & Self-authored novel & Co-authored novel \\
    \hline
    P8 & Web Developer & Customer relationship management software & Novel design interface software \\
    \hline
    P9 & Dancer & Interdisciplinary performance series & Compilation of previous dance work \\
    \hline
    P10 & YouTuber & Videos on personal YouTube Channel & Videos produced for university social media management job \\
    \hline
    P11 & Filmmaker & Wrote, directed, and edited short film & Assistant director and ghostwriter on a film project \\
    \hline
    P12 & Sound Producer & Wrote, produced, and mixed a song & Small sound contributions to commercial sound project \\
    \hline
    P13 & Fiber Artist & Double hoop embroidery work & Fabric collage work \\
    \hline
    P14 & Improv Comedian & Produced and acted in improv comedy show & Made small edits to and acted in a skit \\
    \hline
    P15 & Photographer & Multi-year personal photography project & Commercial photography for a dance company \\
    \hline
    P16 & Cartoonist & Starting a resale print micro-press & Co-director of a comic convention \\
    \hline
    P17 & Illustrator, Graphic Designer & Custom illustrated sticker set for a business & Projects in corporate marketing \\
    \hline
    P18 & Painter, Sculptor & Series of paintings for a personal project & Painting commission \\
    \hline
    P19 & Painter, Glass Artist & Solo art show for senior thesis in university & Commission pieces \\
    \hline
    P20 & Ceramicist & Handmade ceramic piece sold at auction & Oversaw a collaborative paint-by-numbers style project \\
    \hline
    P21 & Costume Designer & Theater play costume design with full creative direction & Theater play with period inspired costumes \\
    \hline
\end{tabularx}
\end{table*}

\subsection{Interview Protocol}

We conducted semi-structured interviews lasting 45–60 minutes, guided by a shared set of questions and thematic prompts while allowing flexibility for participants to reflect on their individual experiences. This approach encouraged rich, situated accounts of ownership while maintaining comparability across interviews. The full interview guide is included in \Cref{app:interview}.

Interviews were structured into two phases. In the \textbf{pre-webtool phase}, participants first provided background information on their creative trajectory, education, and domain of practice. They then reflected on two creative products selected in advance—one associated with high ownership and one with low ownership—explaining the reasoning behind their classifications and the factors that influenced them.

In the \textbf{post-webtool phase}, participants were introduced to the Creative Ownership Webtool, which asked them to evaluate each product across the nine subdimensions of the Person, Process, and System framework, resulting in a numerical value for each project. Finally, participants reflected on the framework outputs, discussing whether the results aligned with their intuitions, which dimensions resonated or felt less relevant, and what aspects of ownership they felt might be missing.

\subsection{Transcript Analysis}

We analyzed interview transcripts using thematic analysis. Each transcript was segmented into meaningful units (quotes or lines), which were then coded based on the core theme or idea expressed. Codes were iteratively refined and collapsed, with similar codes grouped together into broader categories that reflected shared orientations toward ownership. Through repeated reduction, these categories were distilled into a set of central themes that captured the most salient patterns across the dataset.

This bottom-up process enabled us to move from individual statements to collective insights, while preserving the contextual richness of participants’ reflections. Thematic analysis was chosen because it allowed us to account for both recurring concepts across participants and nuanced differences in how ownership was experienced across creative domains.

\section{Results}

\subsection{Quantitative Findings}
\begin{figure}[H]
    \centering 
    \includegraphics[width=\textwidth]{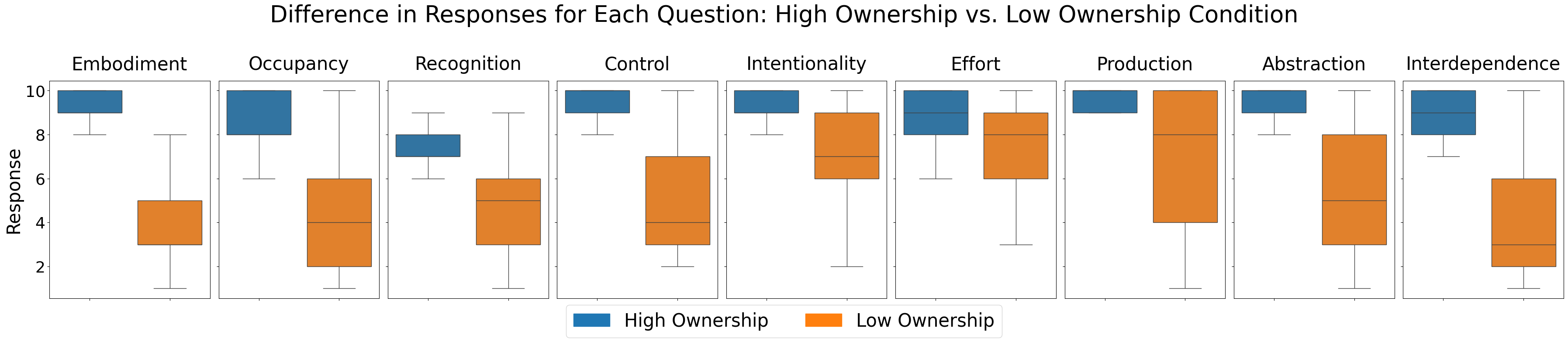} 
    \caption{Difference in Responses for Each Question: High Ownership vs. Low Ownership Condition}
    \label{fig:results}
\end{figure}
Across all nine sub-dimensions of the framework---Embodiment, Occupancy, Recognition, Control, Intentionality, Effort, Production, Abstraction, and Interdependence---participants gave consistently higher ratings for projects they associated with high ownership compared to low ownership (\Cref{fig:results}). This pattern held across the board, suggesting that the framework reliably distinguishes between ownership conditions rather than capturing isolated dimensions.

Responses for low-ownership projects showed substantially greater variance, with wider inter-quartile ranges and more outliers than in the high-ownership condition. Whereas ratings for high-ownership projects clustered tightly at the upper end of the scale, low-ownership responses spanned nearly the full range, from near zero to moderately high values. This indicates that while participants converge on what constitutes high ownership, experiences of low ownership are more heterogeneous, reflecting different ways ownership may be diminished (e.g., limited control, lack of recognition, or minimal effort).

Overall, these results demonstrate both the coverage and diagnostic power of the framework: all nine sub-dimensions shifted between conditions, and the variance patterns in the low ownership condition surfaced the diverse ways participants experience reduced ownership.

\subsection{Pre-Framework}

\textbf{Person: Embodiment}---Embodiment of values, personality, and identity was repeatedly cited by participants as a strong reason why they feel creative ownership over their work. P4 (nonfiction writer) cited that they chose the work because it was both crafted in their signature style, and was an emotional piece written about their mother. P19 (painter, glass artist) chose a piece that was an exploration of body and memory: "It was a lot of looking through and reflecting what I was thinking." Participants used a variety of words to get this message across: self-indulgence, passion, obsession, vulnerability. Being able to engage in their own explorations, share their backgrounds and experiences, and, in the words of one participant, "imbue more of [themselves]" (P9, dancer), was key across the study. Additionally, beyond their sentiments about the works being reflective of themselves and their experiences, several participants cited their sense of ownership stemming from the impact it had on the people in their lives: "I actually felt more attached to the work because people I love felt connected to the work…" (P9, dancer); "The real art in itself, to me, was actually in the community that was born from it…" (P19, painter, glass artist). Beyond personal identity, creative identity also plays a role. Participants felt that when the work reflected their "signature style" (P4, nonfiction writer) or "distinctive mark" (P8, web developer), they had a stronger sense of creative ownership. Conversely, works that were not reflective of their style---even if they had been previously---were associated with lower ownership feelings: "…it wasn't the style I was really working with anymore and I just wanted to sell that piece and move onward." (P18, painter, sculptor). 

\textbf{Person: Occupancy}---The amount of time spent working on the final project seems to have an impact on participants’ sense of ownership. Several participants reported that they had either extensive or unlimited time to spend on their high ownership projects: "…all my time was dedicated to making that, you know? It's like I wasn't thinking about anything else. That's not often the case" (P19, painter, glass artist). Conversely, shorter time frames were associated with low ownership projects: "Most marketing materials have really quick turnaround dates, and I maybe get one day or two days to work on each graphic" (P17, illustrator, graphic designer). 

\textbf{Person: Recognition}---Only three participants mentioned external recognition in any explicit format: P21 (costume designer) noted that it does indeed contribute to their sense of creative ownership. P18 (painter, sculptor) mentioned that there is sometimes a disconnect between the amount of effort that goes into a and the pieces that people relate to the most, and P20 (ceramicist) mentioned that they like to talk to the people that purchase their pieces about what they liked most about them so that they can try to replicate those techniques. Additionally, two participants (P9, dancer; P19, painter, glass artist) noted that the "commodification" of their work---which is a particular type of external recognition---often makes them feel detached from the project. Others also expressed this sentiment, but less explicitly. Also adjacent to the idea of commodification is that of credit. P11 (filmmaker) and P12 (sound producer) noted that projects where they are not credited or given royalties are associated with lower ownership. However, credit does not necessarily translate to high ownership, as P19 (painter, glass artist) noted that although they do maintain copyright over commissions that they create, "it’s not the type of work that [they] would create."

\textbf{Process: Control}---Control was the most cited aspect of creative ownership, with somewhere between a quarter and a third of all comments at this stage of the interview were about having control during the creative process. Being the one to make important decisions and choose the solutions that felt right to them were important to most participants. P6 (video game developer) stated that being in charge of the creative-decision making is vital to their sense of creative ownership: "I had to make a lot of decisions… it was the first time that I had to make those kinds of executive calls. … That’s the highest ownership I’ve felt over a creative project." In addition to this pivotal role, there are many other ways that control can show up in relation to ownership. For P21 (costume designer), this looks like having control over every single element of making a costume come to life: "So that's when you put the time, and you design everything from scratch.  … Those are the ones where I feel like I can claim ownership." For P18 (painter, sculptor), it looks like having clients giving them free reign on commissions: "People have kind of let me go for it. After an initial concept is sketched out." Beyond being given a sense of agency or being put in charge of decision-making, having a lack of restrictions or guidelines is also important to feeling in control and thereby ownership. Having too many parameters (P9, dancer), being beholden to someone else (P19, painter, glass artist), or having to follow a certain script (P11, filmmaker), were all associated with a diminished sense of creative ownership. In the most extreme cases, it can lead to alienation from the work: "I have no control over it, because I’m technically just helping them out. What they do with my scenes, what they do with the dialogue, how it is shot, edited, interpreted---it has nothing to do with me. I have no… not even a little bit of ownership on it" (P11, filmmaker). While all of these external constraints were reported to be associated with a diminished sense of ownership, artistic constraints and self-imposed limitations (P15, photographer), as well as limitations of the material properties and the participant’s abilities (P19, painter, glass artist), were not seen as lowering creative ownership in the same way.

\textbf{Process: Intentionality}---Only one participant directly mentioned the term intentionality, but a few participants reported that whether or not they were able to work on the project from start to finish (a sense of continuity perhaps) was important to their sense of ownership. For instance, P2’s (ukulelist, singer) high ownership project was a song that they worked on from the very beginning to the very end, while their low ownership project involved a song that was already in production, they spent a short amount of time on, and were not even notified of when it was released. P5 (architect) and P8 (web developed) also noted that being part of all of the stages of production were key to a high sense of creative ownership. While continuity is distinct from control or intentionality, it can still shape one’s capacity to make intentional creative decisions, particularly when involvement is limited to a part rather than the whole project. The one participant who did directly reference intentionality did so more in terms of the medium they work with: "We're still digging up shards of pottery from hundreds and thousands of years ago; once you fire something, it doesn't go away. It's hard as rock. So you really want to be sure and confident and intentional when you make something out of clay and fire it, because it can't be undone" (P20, ceramicist). 

\textbf{Process: Effort}---The amount of effort that went into a project was also seldom referenced, with P12 (sound producer) noting that their work often does not feel like work because they find the project "flowing out of [them]". On the other hand, P20 (ceramicist) views the final product as a "representation of [their] labor." 

\textbf{System: Production}---While none of the participants explicitly referred to physically working with the final product, this notion was alluded to by several participants: "I wrote the entire script, recorded the whole thing with a friend over 2 or 3 days, and then posted it" (P10, youtuber) and "It's something that I've written, directed, and edited…I have complete ownership of everything in it" (P11, filmmaker). Similarly, being physically removed from the final product was reported a reason for diminished ownership: "The dress, which is the thing that cuts more [of] your attention, was already designed, by a previous designer that worked here 10 years ago. So how am I gonna put these dresses on my website and say that I designed it?" (P21, costume designer). This degree of separation was also noted by P5 (architect): "To show your professor, your client, your contractor---they need to see the drawing. Because you can’t express in words what you are designing. So, if the drawing is not yours, then the sense of ownership starts to shift, you know?"

\textbf{System: Abstraction}---What participants did refer to more directly, however, is the idea of conceptually working closely with the final product. Most stated that being the driver of the concept or "creative vision" behind a project was crucial to their ownership feelings: "I feel a great sense of ownership, because I was given the job of framing the project narrative and the project concept" (P5, architect). However, more often than not, it was stated in the converse---working on a project that wasn’t aligned with their personal creative vision was associated with lower feelings of creative ownership. P20 (ceramicist) stated that they felt rather apathetic towards a project that was not a result of their vision: "It wasn't personal. I was helping someone else. Um, I was a hired hand to help someone else communicate their message for them." P8 (web developer) also echoed this sentiment: "The thing that motivated that project definitely made me feel like I cared less about it and I don’t think about it now, and I don’t feel proud of it, and I don’t show it off." Even in situations where one had some say in shaping the concept, this minimal involvement was not necessarily enough to overcome a lack of ownership: "I didn’t necessarily participate in writing. I was there to make some edits and suggest alternatives, but for the most part, I was there to perform the sketches, to read what the writers wrote down. So, in that sense, I felt less ownership. I didn’t really create the content" (P14, improv comedian). Related to this sentiment is the originality of the concept, with participants reporting higher ownership for projects they felt were the result of their original ideas: "I've created it, and it's original from me… I feel like I have a strong ownership over it because I'm giving it life, versus it's a visual that's just existing" (P17, illustrator, graphic designer).

\textbf{System: Interdependence}---Finally, whether the work was created independently or in a team setting appears to influence participants’ sense of ownership, with independently created works generally being associated with higher ownership, and works created in group settings being associated with lower ownership. As some participants stated, this could be the result of the fact that independent endeavours put all the onus on the individual by their very nature: "I did everything… I guess as a consequence, I felt like it was objectively clear that that was my product" (P8, web developer). An equal but opposite impact was observed in participants who were working in larger teams: "It's the team that's producing it and you don't feel like you really own anything, okay? You've done 5\% of what was to be done" (P5, architect). In one scenario, the participant was only working with one other person but felt low ownership as their partner carried most of the project by themselves: "And the reason I feel less ownership with that project is because really, [my partner] began the project on her own and did all of that background work before I came to the project" (P7, novelist). In contrast to this, being the "conductor" (P19, painter, glass artist), "orchestrating" the project (P6, video game developer), or having the final "responsibility" (P9, dancer) was reported to be associated with high ownership feelings over a project.

\subsection{Post-Framework}

\textbf{Person: Embodiment}---The sentiments highlighting the importance of embodiment largely paralleled those expressed prior to the participants viewing the framework. Participants stated that it was important to them that their work reflected their "value system" (P5, architect), "emotional experience in [their] lived feelings" (P2, ukulelist, singer), and that it was a "labor of love" (P16, cartoonist). However, some participants felt that embodiment was not central to their work. For some, embodiment just wasn’t really a consideration in their creative practice, like for P18 (painter, sculptor): "I’m coming from a graphic and industrial design background, so I’m not carrying as much story or history in that respect. It’s more about exploring and highlighting things I see in my day-to-day. But it’s not as much of a personal story." However, others strictly wanted to avoid "imposing [their] values" on a project: "…artistic values a little, maybe, but certainly not ideas about how the world should be" (P15, photographer). For others, values took the form of their ability to make broader quality judgments rather than something tied to a personal narrative: "It just matters to me that I was able to be the arbiter on consistency and quality" (P8, web developer). There almost appears to be a divide between "process-focused" (P18, painter, sculptor) and "person-focused" (P3, dancer) creatives. 

\textbf{Person: Occupancy}---Although the majority of participants noted that how much they thought about the final product was not particularly indicative of their ownership feelings, a few reported that it is important to their creative process. P3 (dancer) notes that thinking about the product can open up new ways of extending the project in new directions. P20 (ceramicist) expressed a similar notion, stating that they deliberate whether or not they’re thinking about the work in a productive manner. They noted that this could be a result of the medium they work with: "I think that maybe because the ceramic medium is so permanent, I often find myself thinking about the finality and the permanence of my work" (P20, ceramicist). There might be other confounding factors at play. As P4 (nonfiction writer) brought up, works that they more readily talk about or share with others are also the works that they spend more time thinking about just because they have had the opportunity to have more extensive conversations and develop dialogue around those pieces. 

\textbf{Person: Recognition}---While external recognition was seldom mentioned in the pre-framework phase, several participants stated that it is incredibly important to their sense of ownership over a piece of work. As P11 (filmmaker) put concisely --- "External recognition is a factor, because the more you get recognized for your work, the more you feel that it's yours." Recognition seems to especially play a big role in fields where the work is made for high visibility: "External recognition, I think, sticks out just as a part of what I do, just because a lot of my choreography…could be recognized by someone or an organization" (P3, dancer). P5 (architect) highlighted how this is taken to an extreme in the field of architecture, where acknowledgement, recognition, and patronage play a crucial role in how the participant thinks about their work --- "It stops at recognition, and we don't really feel that sort of need for ownership." However, not all participants felt that recognition was vital to their work, and for some, it wasn’t really a factor at all --- "I’m not thinking a lot about what other people will think of it, or how it will be received. It’s more that it’s something in me that needs to be made, and I’m just focused on making it" (P13, fiber artist).

\textbf{Process: Control}---Similar to the pre-framework phase, participants stated that having greater control was associated with a higher sense of creative ownership and vice versa: "obviously you would have more ownership because there's less friction" (P8, web developer); "How much control did you have over the creative decisions? I would say that's the most important" (P11, filmmaker); "The more they feel like orders at that point, then the less I feel like I own a specific project" (P17, illustrator, graphic designer). However, in addition to these general observations, some participants noted that the source of restrictions or constraints may play a role: one participant noted that the source of their constraints was their job, "Why were there restrictions?…if there were restrictions, did they come from you, or was it because someone else put restrictions on you?… Self-inflicted or external, I guess, would be the binary choices. But I'm sure there's a gray area" (P10, youtuber). P15 (photographer) also elaborated on this, stating that while "creative constraints" that might come from the medium (such as not being able to take photos in restricted areas) or from themselves wouldn’t even make them consider their ownership any differently, "collaborative constraints" that are imposed on them by other people would change their sense of ownership. In some cases, participants were even "intentional about being unintentional" (P12, sound producer and P15, photographer). 

\textbf{Process: Intentionality}---However, there seem to be times when material constraints can indeed shift ownership feelings, especially when control, intentionality, and creative vision all lie at an intersection: "I lose ownership points there, because I'm limited by this specific tool even if I have a specific vision" (P4, nonfiction writer); "I wrote everything that I wanted to, I planned everything the way that I wanted it to be. But when I went to shoot, and I started facing challenges, I realized I don't have enough time, enough budget, and the crew is not experienced enough. So then, your idea of making the film itself changes" (P11, filmmaker). While these constraints are not necessarily coming from employers or collaborators, they limit the participants’ intentionality in relation to executing their creative vision. Even though the majority of participants stated that intentionality doesn’t play a role in their conceptions of ownership as it is "a given" (P5, architect) and that "everything is intentional" (P17, illustrator, graphic designer), these cases showcase that intentionality can indeed play a role in ownership sentiments, especially when the ability to be intentional is taken away. 

\textbf{Process: Effort}---Similarly, most participants said that the amount of physical and mental effort they put into a project doesn’t really impact their ownership feelings, with one participant stating that they "don’t mind that it was easier" (P2, ukulelist, singer) and another stating that "art flows out of [them]" (P12, sound producer). Even still, a few participants noted that it was important: "How much physical and or mental effort. It was definitely helpful" (P9, dancer). Overall, the process section appears to be rather divisive in terms of how participants talked about their ownership feelings with some attributing it as the most important section, while others didn’t even really consider it prior to the framework: "I think the process section of that? In terms of what control I had, how intentional I was, and how much effort I put into it. That's probably most impactful" (P1, social media); "And I didn't think at first that [the process] would be considered part of ownership" (P3, dancer).  This divide could tie back into the person-focused vs. process-focused practices as discussed in the previous section. 

\textbf{System: Production and Abstraction}---Although the ideas of concept and creative vision were repeatedly discussed in the pre-framework phase, there was little follow up in the post-framework phase apart from reiteration that driving the concept and owning the vision are important to a sense of ownership --- "I have the ability to create content, and that's what matters" (P4, nonfiction writer). Conversely, while the notion of physically working closely with the final product was hardly referenced in the pre-framework phase, many participants took note of it in this phase, particularly in relation to conceptually working closely with the final product. The key observation is that this differentiation is highly medium dependent: "I think the physical aspect of it is not applicable to every artist" (P4, nonfiction writer); "I don't fully understand physically working with the final product; I'm sure you're interviewing visual artists, people who do paint on paper and stuff, and it's just not quite as tactile to me" (P2, ukulelist, singer). And indeed, a painter did note that this differentiation was key to their understanding --- "I think talking about the physical versus the sort of conceptual. That really resonates with me…I'm primarily a painter and within my work I pack quite a deal of not only, you know, sort of physical work, but then conceptual work" (P19, painter, glass artist). The tangibility of a medium seems to play a role in whether participants are thinking about how closely they are physically working with the final product and what that means in terms of their ownership. 

\textbf{System: Interdependence}---Finally, in terms of independence, a few participants did reiterate that working independently was associated with a higher sense of ownership, while being a small part of a large team was associated with a lower sense of ownership. However, a new observation that arose was that many participants cited that the nature and context of the project was important to them, with several citing that projects they completed for a job were associated with a lower sense of creative ownership: "It's not creative anymore. It's a routine. It's like any other job" (P11, filmmaker); "It was a job. And when something is a job, you just have to get it done" (P10, youtuber); "In a professional setting you are not working on a brief that you make; you address somebody's needs, and you're actually problem solving" (P5, architect). 

\textbf{Missing Dimensions}---When asked if they felt that there were any dimensions missing from the framework, participants gave a range of answers that could generally be mapped onto one of the existing dimensions or conceptualized as a sub-dimension. Several participants wanted more questions about the ideation process (P4, nonfiction writer), the source of ideas (P20, ceramicist), and the novelty or originality of ideas (P14, improv comedian), which could all be thought of as relevant to the concept behind a piece of work. Many participants thought that it was important to consider how closely the final product aligned with their initial conceptions (P7, novelist; P8, web developer; P11, filmmaker), "almost like a success-type question" (P3, dancer). This idea can be thought of as an aspect of intentionality --- as P11 (filmmaker) stated, "Did your intentions translate into the final work?" Some participants reported feeling greater ownership over projects that were outside their comfort zone (P4, nonfiction writer) or more challenging tasks (P9, dancer). This difficulty or challenge metric can be thought of as factoring into the amount of physical or mental effort exerted by a participant. Some other ideas that came up were creative leadership (P6, video game developer), opportunity to share the work in desired venues (P15, photographer), and purpose and vulnerability (P9, dancer), which can be thought of in terms of control, external recognition, and embodiment respectively.

In addition to these comments, some participants brought up feelings that aren’t quite creative ownership, but appear to be adjacent in some way. P2 (ukulelist, singer) reported feeling a "creative attachment" to a piece, even though they didn’t feel any ownership over it --- "A little bit of my heart and the soul is in this thing, even though it doesn't have anything to do with me otherwise." P4 (nonfiction writer) reported a similar sentiment but used the term pride instead --- "That sense of proudness doesn’t really have anything to do with how much I feel ownership about it, at least not directly." As one participant put simply, "Did I love it?" (P3, dancer).

\section{Discussion}

This study had two aims: to surface mechanisms that shape felt creative ownership and to evaluate whether a structured framework elicits dimensions that participants do not spontaneously report. Pre-framework interviews concentrated on \emph{Embodiment}, \emph{Control}, and \emph{Abstraction}. With the framework in view, attention distributed across all nine dimensions. Quantitatively, high-ownership cases exhibited higher overall scores, whereas low-ownership cases showed greater dispersion. Taken together, these patterns indicate that the framework broadens the analytic space of ownership and supports the capture of heterogeneous routes to ownership, particularly in low-ownership contexts.

A key contribution is comparability. “Ownership” is frequently treated as a unitary construct, obscuring what prior work has manipulated or measured. Naming dimensions enables researchers to specify, for example, that a study manipulates \emph{Interdependence} and \emph{Control} while measuring \emph{Recognition} as a moderator. This improves study design (clearer manipulations), reporting (coverage profiles rather than a single index), and synthesis (cross-study comparisons). Participants also found the categories legible, and a recurrent split emerged between person-focused and process-focused practices. Employment context further moderated ownership: low-ownership projects were often job-driven, whereas high-ownership projects skewed toward self-initiated work. These findings support modeling ownership as a multi-dimensional profile with moderators rather than a single latent factor.

The framework yields actionable implications for system design. Treating ownership as a first-class experience goal positions each dimension as a design lever. \emph{Control} can be protected by making decision rights explicit, keeping suggestions reversible, and attaching rationales to consequential edits. \emph{Intentionality} can be supported through periodic intent check-ins and visual diffs that surface drift from initial goals. \emph{Recognition} benefits from attribution by default. \emph{Production} and \emph{Abstraction} suggest modality-aware workflows (concept-first versus material-first), and \emph{Interdependence} calls for role visibility and decision traceability in collaborative tools. The aim is not to prescribe features but to make ownership \emph{designable}: systems can be tuned to the ownership profile a context demands.

The tool also provided reflective value. Participants reported that it helped articulate what matters to them and why. Beyond research settings, individuals can use the framework to audit which dimensions drive their own sense of ownership, select AI tools that respect those priorities (e.g., suggestion-only assistance for high-\emph{Control} creators), and mediate collaboration by visualizing divergent ownership profiles when teammates disagree about contribution and credit. A potential risk is profile drift under sustained high-automation use (e.g., declines in perceived \emph{Effort} or \emph{Control}). Because the framework is lightweight, it can function as a periodic check-in to track such changes and recommend countermeasures (e.g., adding decision checkpoints or narrowing automation scope).

Methodologically, we recommend reporting an ownership \emph{profile} rather than a single score and explicitly stating construct boundaries. A brief “ownership design card” in Methods---specifying manipulated versus measured dimensions, expected moderators (e.g., medium tangibility, employment context), and anticipated trade-offs---would improve interpretability and comparability. More broadly, the framework shifts inquiry from “how much did this feel like yours?” to “why did it feel like yours, and along which levers could we intervene?” That shift makes results more comparable across studies and, crucially, renders ownership amenable to intentional support in AI-supported creative practice.

\section{Limitations} 

The nine-dimension organization is defensible but not unique; alternative taxonomies would emphasize different cuts. Participants did not spontaneously propose wholly new dimensions, but we do not claim exhaustiveness. We also make theory-laden boundary choices---treating \emph{Control} as a dimension of ownership and keeping pride/enjoyment adjacent rather than inside the construct---which warrant continued debate. Finally, participants selected their own high and low ownership projects. This introduces salience and survivorship biases (forgettable, very low-effort artifacts are under-sampled). Cross-domain sampling improves coverage at the cost of within-domain nuance. These trade-offs delimit inference but are appropriate for establishing coverage.

\section{Future Work}

Rather than treating dimensions independently, future work should examine their interactions. For example, we repeatedly observed \emph{Control} entangled with both \emph{Intentionality} (ability to execute one’s vision) and \emph{Embodiment} (expressing identity/values). A second direction is within-domain depth: interviewing a cohort within a single field (e.g., 20 choreographers, or 20 ceramicists) can surface nuances and interaction patterns obscured by cross-field sampling. Finally, education contexts invite domain-specific hypotheses: in creative writing, \emph{Embodiment} may be primary, whereas in mathematics instructors may prioritize \emph{Intentionality}. The multidimensional nature of this framework sheds light on these differences when it comes to how we think about evaluating the impacts of a system.

\section{Conclusion}

We introduced a nine-subdimension framework for creative ownership and an interactive web tool, and evaluated both through interviews with 21 creative professionals. The framework broadened what participants discussed beyond a few familiar themes and separated high- from low-ownership cases while revealing heterogeneity in how ownership diminishes.

For HCI, the immediate use is practical: report ownership as a \emph{profile} rather than a single score, state construct boundaries, and use the dimensions as design levers (e.g., decision rights for \emph{Control}, intent alignment for \emph{Intentionality}, attribution for \emph{Recognition}, modality-aware workflows for \emph{Production}/\emph{Abstraction}, and role clarity for \emph{Interdependence}).

Next steps include testing interactions among dimensions, running within-domain and longitudinal studies, and using the framework as an outcome measure when evaluating AI-in-the-loop tools. Adoption of the framework and tool should make ownership findings more comparable across studies and help designers build systems that support, rather than erode, felt ownership.

\pagebreak

\nocite{*}
\bibliographystyle{ACM-Reference-Format}
\bibliography{references}


\begin{thebibliography}{47}


\ifx \showCODEN    \undefined \def \showCODEN     #1{\unskip}     \fi
\ifx \showISBNx    \undefined \def \showISBNx     #1{\unskip}     \fi
\ifx \showISBNxiii \undefined \def \showISBNxiii  #1{\unskip}     \fi
\ifx \showISSN     \undefined \def \showISSN      #1{\unskip}     \fi
\ifx \showLCCN     \undefined \def \showLCCN      #1{\unskip}     \fi
\ifx \shownote     \undefined \def \shownote      #1{#1}          \fi
\ifx \showarticletitle \undefined \def \showarticletitle #1{#1}   \fi
\ifx \showURL      \undefined \def \showURL       {\relax}        \fi
\providecommand\bibfield[2]{#2}
\providecommand\bibinfo[2]{#2}
\providecommand\natexlab[1]{#1}
\providecommand\showeprint[2][]{arXiv:#2}

\bibitem[Aikhenvald(2012)]%
        {Aikhenvald2012}
\bibfield{author}{\bibinfo{person}{Alexandra~Y. Aikhenvald}.} \bibinfo{year}{2012}\natexlab{}.
\newblock \showarticletitle{Possession and Ownership: A Cross-Linguistic Perspective}.
\newblock In \bibinfo{booktitle}{\emph{Possession and Ownership}}, \bibfield{editor}{\bibinfo{person}{Alexandra~Y. Aikhenvald} {and} \bibinfo{person}{R.~M.~W. Dixon}} (Eds.). \bibinfo{publisher}{Oxford University Press}, \bibinfo{pages}{1--64}.
\newblock


\bibitem[{Berkeley International Office}(2025)]%
        {BerkeleyCopyrights}
\bibfield{author}{\bibinfo{person}{{Berkeley International Office}}.} \bibinfo{year}{2025}\natexlab{}.
\newblock \bibinfo{booktitle}{\emph{Copyrights: Protecting Creators and Their Creative Expressions}}.
\newblock
\urldef\tempurl%
\url{https://internationaloffice.berkeley.edu/students/intellectual-property-guide-uc-berkeley-graduate-students/copyrights-protecting-creators}
\showURL{%
\tempurl}
\newblock
\shownote{Accessed: 2025-03-19}.


\bibitem[Biermann et~al\mbox{.}(2022)]%
        {biermann2022fromtool}
\bibfield{author}{\bibinfo{person}{O.~C. Biermann} {et~al\mbox{.}}} \bibinfo{year}{2022}\natexlab{}.
\newblock \showarticletitle{From Tool to Companion: Storywriters Want AI Writers to Respect Their Personal Values and Writing Strategies}. In \bibinfo{booktitle}{\emph{Proceedings of the Designing Interactive Systems Conference}}. \bibinfo{address}{Virtual Event, Australia}, \bibinfo{pages}{1209--1227}.
\newblock
\href{https://doi.org/10.1145/3532106.3533506}{doi:\nolinkurl{10.1145/3532106.3533506}}


\bibitem[Boyer(2023)]%
        {Boyer2023}
\bibfield{author}{\bibinfo{person}{Pascal Boyer}.} \bibinfo{year}{2023}\natexlab{}.
\newblock \showarticletitle{Ownership Psychology as a Cognitive Adaptation: A Minimalist Model}.
\newblock \bibinfo{journal}{\emph{Behavioral and Brain Sciences}}  \bibinfo{volume}{46} (\bibinfo{year}{2023}), \bibinfo{pages}{e323}.
\newblock
\href{https://doi.org/10.1017/S0140525X22002527}{doi:\nolinkurl{10.1017/S0140525X22002527}}


\bibitem[Brunneder and Dholakia(2018)]%
        {brunneder2018self}
\bibfield{author}{\bibinfo{person}{J. Brunneder} {and} \bibinfo{person}{U. Dholakia}.} \bibinfo{year}{2018}\natexlab{}.
\newblock \showarticletitle{The self-creation effect: making a product supports its mindful consumption and the consumer’s well-being}.
\newblock \bibinfo{journal}{\emph{Marketing Letters}} \bibinfo{volume}{29}, \bibinfo{number}{3} (\bibinfo{date}{Sept.} \bibinfo{year}{2018}), \bibinfo{pages}{377--389}.
\newblock
\href{https://doi.org/10.1007/s11002-018-9465-6}{doi:\nolinkurl{10.1007/s11002-018-9465-6}}


\bibitem[Buchem(2012)]%
        {Buchem2012}
\bibfield{author}{\bibinfo{person}{Ilona Buchem}.} \bibinfo{year}{2012}\natexlab{}.
\newblock \showarticletitle{Psychological Ownership and Personal Learning Environments: Do Sense of Ownership and Control Really Matter?}. In \bibinfo{booktitle}{\emph{Proceedings of the PLE Conference 2012}}. \bibinfo{address}{Aveiro, Portugal}.
\newblock
\urldef\tempurl%
\url{https://www.semanticscholar.org/paper/Psychological-Ownership-and-Personal-Learning-Do-of-Buchem/b37bf6db5a3ebb806692de2f47a907f241e7dbfc}
\showURL{%
\tempurl}


\bibitem[Cherry and Latulipe(2014)]%
        {cherry2014quantifying}
\bibfield{author}{\bibinfo{person}{E. Cherry} {and} \bibinfo{person}{C. Latulipe}.} \bibinfo{year}{2014}\natexlab{}.
\newblock \showarticletitle{Quantifying the Creativity Support of Digital Tools through the Creativity Support Index}.
\newblock \bibinfo{journal}{\emph{ACM Transactions on Computer-Human Interaction}} \bibinfo{volume}{21}, \bibinfo{number}{4} (\bibinfo{date}{Aug.} \bibinfo{year}{2014}), \bibinfo{pages}{1--25}.
\newblock
\href{https://doi.org/10.1145/2617588}{doi:\nolinkurl{10.1145/2617588}}


\bibitem[Draxler et~al\mbox{.}(2024)]%
        {draxler2024aighostwriter}
\bibfield{author}{\bibinfo{person}{F. Draxler} {et~al\mbox{.}}} \bibinfo{year}{2024}\natexlab{}.
\newblock \showarticletitle{The AI Ghostwriter Effect: When Users do not Perceive Ownership of AI-Generated Text but Self-Declare as Authors}.
\newblock \bibinfo{journal}{\emph{ACM Transactions on Computer-Human Interaction}} \bibinfo{volume}{31}, \bibinfo{number}{2} (\bibinfo{date}{April} \bibinfo{year}{2024}), \bibinfo{pages}{1--40}.
\newblock
\href{https://doi.org/10.1145/3637875}{doi:\nolinkurl{10.1145/3637875}}


\bibitem[Fox(2023)]%
        {Fox2023}
\bibfield{author}{\bibinfo{person}{Katie Fox}.} \bibinfo{year}{2023}\natexlab{}.
\newblock \showarticletitle{Struggle: What is it Good For?}
\newblock \bibinfo{journal}{\emph{Georgia Council of Teachers of Mathematics}} (\bibinfo{year}{2023}).
\newblock
\urldef\tempurl%
\url{https://gctm.org/Struggle-What-is-it-Good-For}
\showURL{%
\tempurl}


\bibitem[Furby(1978)]%
        {Furby1978}
\bibfield{author}{\bibinfo{person}{Lita Furby}.} \bibinfo{year}{1978}\natexlab{}.
\newblock \showarticletitle{Possession in Humans: An Exploratory Study of Its Meaning and Motivation}.
\newblock \bibinfo{journal}{\emph{Social Behavior and Personality: an international journal}} \bibinfo{volume}{6}, \bibinfo{number}{1} (\bibinfo{date}{Jan.} \bibinfo{year}{1978}), \bibinfo{pages}{49--65}.
\newblock
\href{https://doi.org/10.2224/sbp.1978.6.1.49}{doi:\nolinkurl{10.2224/sbp.1978.6.1.49}}


\bibitem[Guo et~al\mbox{.}(2024)]%
        {guo2024exploring}
\bibfield{author}{\bibinfo{person}{A. Guo} {et~al\mbox{.}}} \bibinfo{year}{2024}\natexlab{}.
\newblock \showarticletitle{Exploring the Impact of AI Value Alignment in Collaborative Ideation: Effects on Perception, Ownership, and Output}. In \bibinfo{booktitle}{\emph{Extended Abstracts of the CHI Conference on Human Factors in Computing Systems}}. \bibinfo{address}{Honolulu, HI, USA}, \bibinfo{pages}{1--11}.
\newblock
\href{https://doi.org/10.1145/3613905.3650892}{doi:\nolinkurl{10.1145/3613905.3650892}}


\bibitem[Guyer and Horstmann(2015)]%
        {GuyerHorstmann2015}
\bibfield{author}{\bibinfo{person}{Paul Guyer} {and} \bibinfo{person}{Rolf-Peter Horstmann}.} \bibinfo{year}{2015}\natexlab{}.
\newblock \showarticletitle{Idealism}.
\newblock In \bibinfo{booktitle}{\emph{The Stanford Encyclopedia of Philosophy}}, \bibfield{editor}{\bibinfo{person}{Edward~N. Zalta}} (Ed.). \bibinfo{publisher}{Metaphysics Research Lab, Stanford University}.
\newblock
\newblock
\shownote{Available at: \url{https://plato.stanford.edu/archives/fall2015/entries/idealism/}}.


\bibitem[He et~al\mbox{.}(2024)]%
        {he2024ai}
\bibfield{author}{\bibinfo{person}{J. He} {et~al\mbox{.}}} \bibinfo{year}{2024}\natexlab{}.
\newblock \showarticletitle{AI and the Future of Collaborative Work: Group Ideation with an LLM in a Virtual Canvas}. In \bibinfo{booktitle}{\emph{Proceedings of the 3rd Annual Meeting of the Symposium on Human-Computer Interaction for Work}}. \bibinfo{address}{Newcastle upon Tyne, United Kingdom}, \bibinfo{pages}{1--14}.
\newblock
\href{https://doi.org/10.1145/3663384.3663398}{doi:\nolinkurl{10.1145/3663384.3663398}}


\bibitem[Honoré(1961)]%
        {Honore1961}
\bibfield{author}{\bibinfo{person}{Tony Honoré}.} \bibinfo{year}{1961}\natexlab{}.
\newblock \showarticletitle{Ownership}.
\newblock In \bibinfo{booktitle}{\emph{Oxford Essays in Jurisprudence: A Collaborative Work}}, \bibfield{editor}{\bibinfo{person}{Anthony~Gordon Guest}} (Ed.). \bibinfo{publisher}{Oxford University Press}, \bibinfo{address}{New York}, \bibinfo{pages}{107--147}.
\newblock


\bibitem[Joshi and Vogel(2025)]%
        {joshi2025writing}
\bibfield{author}{\bibinfo{person}{N. Joshi} {and} \bibinfo{person}{D. Vogel}.} \bibinfo{year}{2025}\natexlab{}.
\newblock \showarticletitle{Writing with AI Lowers Psychological Ownership, but Longer Prompts Can Help}. In \bibinfo{booktitle}{\emph{Proceedings of the 7th ACM Conference on Conversational User Interfaces}}. \bibinfo{address}{Waterloo, ON, Canada}, \bibinfo{pages}{1--17}.
\newblock
\href{https://doi.org/10.1145/3719160.3736608}{doi:\nolinkurl{10.1145/3719160.3736608}}


\bibitem[Kadoma et~al\mbox{.}(2024)]%
        {kadoma2024role}
\bibfield{author}{\bibinfo{person}{K. Kadoma} {et~al\mbox{.}}} \bibinfo{year}{2024}\natexlab{}.
\newblock \showarticletitle{The Role of Inclusion, Control, and Ownership in Workplace AI-Mediated Communication}. In \bibinfo{booktitle}{\emph{Proceedings of the CHI Conference on Human Factors in Computing Systems}}. \bibinfo{address}{Honolulu, HI, USA}, \bibinfo{pages}{1--10}.
\newblock
\href{https://doi.org/10.1145/3613904.3642650}{doi:\nolinkurl{10.1145/3613904.3642650}}


\bibitem[Koster et~al\mbox{.}(2015)]%
        {koster2015beliefs}
\bibfield{author}{\bibinfo{person}{R. Koster} {et~al\mbox{.}}} \bibinfo{year}{2015}\natexlab{}.
\newblock \showarticletitle{How beliefs about self-creation inflate value in the human brain}.
\newblock \bibinfo{journal}{\emph{Frontiers in Human Neuroscience}}  \bibinfo{volume}{9} (\bibinfo{date}{Sept.} \bibinfo{year}{2015}).
\newblock
\href{https://doi.org/10.3389/fnhum.2015.00473}{doi:\nolinkurl{10.3389/fnhum.2015.00473}}


\bibitem[Kuzminykh and Cauchard(2020)]%
        {kuzminykh2020bemine}
\bibfield{author}{\bibinfo{person}{A. Kuzminykh} {and} \bibinfo{person}{J.~R. Cauchard}.} \bibinfo{year}{2020}\natexlab{}.
\newblock \showarticletitle{Be Mine: Contextualization of Ownership Research in HCI}. In \bibinfo{booktitle}{\emph{Extended Abstracts of the 2020 CHI Conference on Human Factors in Computing Systems}}. \bibinfo{address}{Honolulu, HI, USA}, \bibinfo{pages}{1--9}.
\newblock
\href{https://doi.org/10.1145/3334480.3383058}{doi:\nolinkurl{10.1145/3334480.3383058}}


\bibitem[Lee(2008)]%
        {Lee2008}
\bibfield{author}{\bibinfo{person}{Rosemary Lee}.} \bibinfo{year}{2008}\natexlab{}.
\newblock \showarticletitle{Aiming for Stewardship and not Ownership}.
\newblock In \bibinfo{booktitle}{\emph{An Introduction to Community Dance Practice}}, \bibfield{editor}{\bibinfo{person}{Diane Amans}} (Ed.). \bibinfo{publisher}{Palgrave Macmillan}, \bibinfo{address}{London}.
\newblock
\showISBNx{9780230551695}


\bibitem[Levene and Friedman(2015)]%
        {Levene2015}
\bibfield{author}{\bibinfo{person}{Mark Levene} {and} \bibinfo{person}{Ori Friedman}.} \bibinfo{year}{2015}\natexlab{}.
\newblock \showarticletitle{Creation in Judgments About the Establishment of Ownership}.
\newblock \bibinfo{journal}{\emph{Journal of Experimental Social Psychology}}  \bibinfo{volume}{60} (\bibinfo{date}{Sep.} \bibinfo{year}{2015}), \bibinfo{pages}{103--109}.
\newblock
\href{https://doi.org/10.1016/j.jesp.2015.04.011}{doi:\nolinkurl{10.1016/j.jesp.2015.04.011}}


\bibitem[Liang and Gero(2025)]%
        {liang2025synthia}
\bibfield{author}{\bibinfo{person}{I. Liang} {and} \bibinfo{person}{K.~Ilonka Gero}.} \bibinfo{year}{2025}\natexlab{}.
\newblock \showarticletitle{SYNthia: An Interface Concept for Writing With Large Language Models}. In \bibinfo{booktitle}{\emph{Proceedings of the 2025 Conference on Creativity and Cognition}}. \bibinfo{address}{Virtual, United Kingdom}, \bibinfo{pages}{458--464}.
\newblock
\href{https://doi.org/10.1145/3698061.3734401}{doi:\nolinkurl{10.1145/3698061.3734401}}


\bibitem[Locke(1689)]%
        {Locke1689}
\bibfield{author}{\bibinfo{person}{John Locke}.} \bibinfo{year}{1689}\natexlab{}.
\newblock \bibinfo{booktitle}{\emph{Second Treatise of Government}}.
\newblock
\urldef\tempurl%
\url{https://www.gutenberg.org/files/7370/7370-h/7370-h.htm}
\showURL{%
\tempurl}
\newblock
\shownote{The Project Gutenberg eBook}.


\bibitem[Lovato et~al\mbox{.}(2024)]%
        {lovato2024foregrounding}
\bibfield{author}{\bibinfo{person}{J. Lovato} {et~al\mbox{.}}} \bibinfo{year}{2024}\natexlab{}.
\newblock \showarticletitle{Foregrounding Artist Opinions: A Survey Study on Transparency, Ownership, and Fairness in AI Generative Art}. In \bibinfo{booktitle}{\emph{Proceedings of the AAAI/ACM Conference on AI, Ethics, and Society}}, Vol.~\bibinfo{volume}{7}. \bibinfo{pages}{905--916}.
\newblock
\href{https://doi.org/10.1609/aies.v7i1.31691}{doi:\nolinkurl{10.1609/aies.v7i1.31691}}


\bibitem[Lyu and [Additional~Authors(2023)]%
        {Lyu2023}
\bibfield{author}{\bibinfo{person}{Chenxi Lyu} {and} \bibinfo{person}{if~needed] [Additional~Authors}.} \bibinfo{year}{2023}\natexlab{}.
\newblock \showarticletitle{Travelers’ Psychological Ownership: A Systematic Review and Future Research Agenda}.
\newblock \bibinfo{journal}{\emph{Journal of Travel Research}} \bibinfo{volume}{62}, \bibinfo{number}{8} (\bibinfo{date}{Nov.} \bibinfo{year}{2023}), \bibinfo{pages}{1623--1646}.
\newblock
\href{https://doi.org/10.1177/00472875231151395}{doi:\nolinkurl{10.1177/00472875231151395}}


\bibitem[McClelland(1951)]%
        {McClelland1951}
\bibfield{author}{\bibinfo{person}{David~C. McClelland}.} \bibinfo{year}{1951}\natexlab{}.
\newblock \bibinfo{booktitle}{\emph{Personality}}.
\newblock \bibinfo{publisher}{William Sloane Associates}.
\newblock


\bibitem[McIntyre et~al\mbox{.}(2009)]%
        {McIntyre2009}
\bibfield{author}{\bibinfo{person}{Nancy McIntyre}, \bibinfo{person}{Jon~L. Pierce}, {and} \bibinfo{person}{Sarah L.~F. Rekker}.} \bibinfo{year}{2009}\natexlab{}.
\newblock \showarticletitle{The Relationship of Locus of Control and Motives with Psychological Ownership in Organizations}.
\newblock \bibinfo{journal}{\emph{Journal of Managerial Issues}} \bibinfo{volume}{21}, \bibinfo{number}{3} (\bibinfo{year}{2009}), \bibinfo{pages}{383--401}.
\newblock


\bibitem[Melnyk(2012)]%
        {Melnyk2012}
\bibfield{author}{\bibinfo{person}{Andrew Melnyk}.} \bibinfo{year}{2012}\natexlab{}.
\newblock \showarticletitle{Materialism}.
\newblock \bibinfo{journal}{\emph{WIREs Cognitive Science}} \bibinfo{volume}{3}, \bibinfo{number}{3} (\bibinfo{date}{May} \bibinfo{year}{2012}), \bibinfo{pages}{281--292}.
\newblock
\href{https://doi.org/10.1002/wcs.1174}{doi:\nolinkurl{10.1002/wcs.1174}}


\bibitem[Morewedge(2021)]%
        {Morewedge2021}
\bibfield{author}{\bibinfo{person}{Carey~K. Morewedge}.} \bibinfo{year}{2021}\natexlab{}.
\newblock \showarticletitle{Psychological ownership: Implicit and explicit}.
\newblock \bibinfo{journal}{\emph{Current Opinion in Psychology}}  \bibinfo{volume}{39} (\bibinfo{date}{Jun.} \bibinfo{year}{2021}), \bibinfo{pages}{125--132}.
\newblock
\href{https://doi.org/10.1016/j.copsyc.2020.10.003}{doi:\nolinkurl{10.1016/j.copsyc.2020.10.003}}


\bibitem[Mossoff(2003)]%
        {Mossoff2003}
\bibfield{author}{\bibinfo{person}{Adam Mossoff}.} \bibinfo{year}{2003}\natexlab{}.
\newblock \showarticletitle{What is Property? Putting the Pieces Back Together}.
\newblock \bibinfo{journal}{\emph{SSRN Electronic Journal}} (\bibinfo{year}{2003}).
\newblock
\href{https://doi.org/10.2139/ssrn.438780}{doi:\nolinkurl{10.2139/ssrn.438780}}


\bibitem[Nicholes(2015)]%
        {Nicholes2015}
\bibfield{author}{\bibinfo{person}{Justin Nicholes}.} \bibinfo{year}{2015}\natexlab{}.
\newblock \showarticletitle{Measuring Ownership of Creative Versus Academic Writing: Implications for Interdisciplinary Praxis}.
\newblock \bibinfo{journal}{\emph{Writing in Practice}}  \bibinfo{volume}{3} (\bibinfo{year}{2015}).
\newblock
\urldef\tempurl%
\url{https://www.nawe.co.uk/DB/wip-editions/articles/measuring-ownership-of-creative-versus-academic-writing-implications-for-interdisciplinary-praxis.html}
\showURL{%
\tempurl}


\bibitem[Pierce et~al\mbox{.}(2001)]%
        {Pierce2001}
\bibfield{author}{\bibinfo{person}{Jon~L. Pierce}, \bibinfo{person}{Tatiana Kostova}, {and} \bibinfo{person}{Kurt~T. Dirks}.} \bibinfo{year}{2001}\natexlab{}.
\newblock \showarticletitle{Toward a Theory of Psychological Ownership in Organizations}.
\newblock \bibinfo{journal}{\emph{The Academy of Management Review}} \bibinfo{volume}{26}, \bibinfo{number}{2} (\bibinfo{date}{Apr.} \bibinfo{year}{2001}), \bibinfo{pages}{298}.
\newblock
\href{https://doi.org/10.2307/259124}{doi:\nolinkurl{10.2307/259124}}


\bibitem[Pierce et~al\mbox{.}(2003)]%
        {Pierce2003}
\bibfield{author}{\bibinfo{person}{Jon~L. Pierce}, \bibinfo{person}{Tatiana Kostova}, {and} \bibinfo{person}{Kurt~T. Dirks}.} \bibinfo{year}{2003}\natexlab{}.
\newblock \showarticletitle{The State of Psychological Ownership: Integrating and Extending a Century of Research}.
\newblock \bibinfo{journal}{\emph{Review of General Psychology}} \bibinfo{volume}{7}, \bibinfo{number}{1} (\bibinfo{date}{Mar.} \bibinfo{year}{2003}), \bibinfo{pages}{84--107}.
\newblock
\href{https://doi.org/10.1037/1089-2680.7.1.84}{doi:\nolinkurl{10.1037/1089-2680.7.1.84}}


\bibitem[Pipes(2000)]%
        {Pipes2000}
\bibfield{author}{\bibinfo{person}{Richard Pipes}.} \bibinfo{year}{2000}\natexlab{}.
\newblock \bibinfo{booktitle}{\emph{Property and Freedom}}.
\newblock \bibinfo{publisher}{Vintage Books}.
\newblock


\bibitem[Radin(1982)]%
        {Radin1982}
\bibfield{author}{\bibinfo{person}{Margaret~Jane Radin}.} \bibinfo{year}{1982}\natexlab{}.
\newblock \showarticletitle{Property and Personhood}.
\newblock \bibinfo{journal}{\emph{Stanford Law Review}} \bibinfo{volume}{34}, \bibinfo{number}{5} (\bibinfo{date}{May} \bibinfo{year}{1982}), \bibinfo{pages}{957}.
\newblock
\href{https://doi.org/10.2307/1228541}{doi:\nolinkurl{10.2307/1228541}}


\bibitem[Rochberg-Halton(1980)]%
        {RochbergHalton1980}
\bibfield{author}{\bibinfo{person}{Eugene~W. Rochberg-Halton}.} \bibinfo{year}{1980}\natexlab{}.
\newblock \emph{\bibinfo{title}{Cultural Signs and Urban Adaptation: The Meaning of Cherished Household Possessions}}.
\newblock \bibinfo{thesistype}{Ph.\,D. Dissertation}. \bibinfo{school}{Dissertation Abstracts International Section A: Humanities and Social Sciences}.
\newblock


\bibitem[Rouse(2013)]%
        {Rouse2013}
\bibfield{author}{\bibinfo{person}{Elizabeth~D. Rouse}.} \bibinfo{year}{2013}\natexlab{}.
\newblock \showarticletitle{Kill Your Darlings? Experiencing, Maintaining, and Changing Psychological Ownership in Creative Work}.
\newblock  (\bibinfo{year}{2013}).
\newblock
\urldef\tempurl%
\url{https://www.semanticscholar.org/paper/Kill-your-darlings-Experiencing%2C-maintaining%2C-and-Rouse/9af8484be761e06799824bd66cc3b81b4bea0a73}
\showURL{%
\tempurl}


\bibitem[Sartre(1943)]%
        {Sartre1943}
\bibfield{author}{\bibinfo{person}{Jean-Paul Sartre}.} \bibinfo{year}{1943}\natexlab{}.
\newblock \bibinfo{booktitle}{\emph{Being and Nothingness}}.
\newblock \bibinfo{publisher}{Gallimard}, \bibinfo{address}{Paris}.
\newblock


\bibitem[Smart(1963)]%
        {Smart1963}
\bibfield{author}{\bibinfo{person}{J.J.C. Smart}.} \bibinfo{year}{1963}\natexlab{}.
\newblock \showarticletitle{Materialism}.
\newblock \bibinfo{journal}{\emph{The Journal of Philosophy}} \bibinfo{volume}{60}, \bibinfo{number}{22} (\bibinfo{date}{Oct.} \bibinfo{year}{1963}), \bibinfo{pages}{651}.
\newblock
\href{https://doi.org/10.2307/2023512}{doi:\nolinkurl{10.2307/2023512}}


\bibitem[Unknown({[n.\,d.]})]%
        {SEPProperty}
\bibfield{author}{\bibinfo{person}{Author(s) Unknown}.} \bibinfo{year}{[n.\,d.]}\natexlab{}.
\newblock \showarticletitle{Property}.
\newblock In \bibinfo{booktitle}{\emph{The Stanford Encyclopedia of Philosophy}}, \bibfield{editor}{\bibinfo{person}{Edward~N. Zalta}} (Ed.). \bibinfo{publisher}{Metaphysics Research Lab, Stanford University}.
\newblock
\newblock
\shownote{Available at: \url{https://plato.stanford.edu/entries/property/}}.


\bibitem[Waldron(1985)]%
        {Waldron1985}
\bibfield{author}{\bibinfo{person}{Jeremy Waldron}.} \bibinfo{year}{1985}\natexlab{}.
\newblock \showarticletitle{What Is Private Property?}
\newblock \bibinfo{journal}{\emph{Oxford Journal of Legal Studies}} \bibinfo{volume}{5}, \bibinfo{number}{3} (\bibinfo{year}{1985}), \bibinfo{pages}{313--349}.
\newblock


\bibitem[Wasi et~al\mbox{.}(2024)]%
        {wasi2024llms}
\bibfield{author}{\bibinfo{person}{A.~T. Wasi} {et~al\mbox{.}}} \bibinfo{year}{2024}\natexlab{}.
\newblock \showarticletitle{LLMs as Writing Assistants: Exploring Perspectives on Sense of Ownership and Reasoning}. In \bibinfo{booktitle}{\emph{Proceedings of the Third Workshop on Intelligent and Interactive Writing Assistants}}. \bibinfo{address}{Honolulu, HI, USA}, \bibinfo{pages}{38--42}.
\newblock
\href{https://doi.org/10.1145/3690712.3690723}{doi:\nolinkurl{10.1145/3690712.3690723}}


\bibitem[Weber et~al\mbox{.}(2024)]%
        {weber2024wraiter}
\bibfield{author}{\bibinfo{person}{C.~J. Weber} {et~al\mbox{.}}} \bibinfo{year}{2024}\natexlab{}.
\newblock \showarticletitle{wr-AI-ter: Enhancing Ownership Perception in AI-Driven Script Writing}. In \bibinfo{booktitle}{\emph{Proceedings of the ACM International Conference on Interactive Media Experiences}}. \bibinfo{address}{Stockholm, Sweden}, \bibinfo{pages}{145--156}.
\newblock
\href{https://doi.org/10.1145/3639701.3656325}{doi:\nolinkurl{10.1145/3639701.3656325}}


\bibitem[Weber et~al\mbox{.}(2025)]%
        {weber2025drawing}
\bibfield{author}{\bibinfo{person}{C.~J. Weber} {et~al\mbox{.}}} \bibinfo{year}{2025}\natexlab{}.
\newblock \showarticletitle{Drawing-in-Steps: Supporting Creative Goals through User Engagement via Hierarchical Image Generation}. In \bibinfo{booktitle}{\emph{Proceedings of the 2025 ACM International Conference on Interactive Media Experiences}}. \bibinfo{address}{Niter{\'o}i, Brazil}, \bibinfo{pages}{108--125}.
\newblock
\href{https://doi.org/10.1145/3706370.3727862}{doi:\nolinkurl{10.1145/3706370.3727862}}


\bibitem[White(1959)]%
        {White1959}
\bibfield{author}{\bibinfo{person}{Robert~W. White}.} \bibinfo{year}{1959}\natexlab{}.
\newblock \showarticletitle{Motivation Reconsidered: The Concept of Competence}.
\newblock \bibinfo{journal}{\emph{Psychological Review}} \bibinfo{volume}{66}, \bibinfo{number}{5} (\bibinfo{date}{Sep.} \bibinfo{year}{1959}), \bibinfo{pages}{297--333}.
\newblock
\href{https://doi.org/10.1037/h0040934}{doi:\nolinkurl{10.1037/h0040934}}


\bibitem[Wu et~al\mbox{.}(2023)]%
        {wu2023owndiffusion}
\bibfield{author}{\bibinfo{person}{Y. Wu} {et~al\mbox{.}}} \bibinfo{year}{2023}\natexlab{}.
\newblock \showarticletitle{OwnDiffusion: A Design Pipeline Using Design Generative AI to Preserve Sense of Ownership}. In \bibinfo{booktitle}{\emph{SIGGRAPH Asia 2023 Posters}}. \bibinfo{address}{Sydney, NSW, Australia}, \bibinfo{pages}{1--2}.
\newblock
\href{https://doi.org/10.1145/3610542.3626142}{doi:\nolinkurl{10.1145/3610542.3626142}}


\bibitem[Zhou and Sterman(2023)]%
        {Zhou2023}
\bibfield{author}{\bibinfo{person}{David Zhou} {and} \bibinfo{person}{Sarah Sterman}.} \bibinfo{year}{2023}\natexlab{}.
\newblock \showarticletitle{Creative Struggle: Arguing for the Value of Difficulty in Supporting Ownership and Self-Expression in Creative Writing}. In \bibinfo{booktitle}{\emph{Proceedings of the Second Workshop on Intelligent and Interactive Writing Assistants (In2Writing)}}.
\newblock
\urldef\tempurl%
\url{https://cdn.glitch.global/d058c114-3406-43be-8a3c-d3afff35eda2/paper11_2023.pdf}
\showURL{%
\tempurl}


\bibitem[Émile Durkheim(1957)]%
        {Durkheim1957}
\bibfield{author}{\bibinfo{person}{Émile Durkheim}.} \bibinfo{year}{1957}\natexlab{}.
\newblock \bibinfo{booktitle}{\emph{Professional Ethics and Civic Morals}}.
\newblock \bibinfo{publisher}{Routledge \& Kegan Paul}, \bibinfo{address}{London}.
\newblock


\end{thebibliography}

\appendix

\section{Demonstrating the Difference Between Process and System}\label{sec:process-vs-system}

This example will show that the process dimensions (control, intentionality, and effort) and the system dimensions (production, abstraction, and interdependence) are not always necessarily in alignment, even if it may appear to be so more generally. Imagine an individual who just purchased a paint-by-numbers kit and is completing the project at home. This individual does not have much control or intentionality over the creative decisions impacting the final painting, and may have put in little to moderate effort. In line with this, their score for conceptualization would be low as well as they did not directly conceive of the final outcome. However, their score for production would be very high as they directly worked with the final material outcome of the painting. Therefore, while process dimensions and system dimensions may be positively correlated in some scenarios, this is not always necessarily the case. 

\section{Example Applications of the Framework}\label{sec:examples}

\subsection{The Interior Designer}

The interior designer, representing commissioned creatives more broadly, typically reflects the client's preferences and values in their design choices even if the client may expect some aesthetic input from the designer. The designer invests significant time and receives moderate recognition from the client, firm, and potential guests. They have substantial control over the process and exert effort, but the final intentionality of the space depends on the client. Although they work closely with the final product and conceptualize the design, they collaborate with the client and other stakeholders to execute the vision, with their ownership primarily stemming from enacting the creative process.

\subsection{The Community Arts Practitioner}

The community arts practitioner might be an individual who has a grand creative vision and loops in their community to make it happen. Their ideas likely embody their values along with those of their community, they likely spend a lot of time thinking about the idea, and their community recognizes them for this piece of (generally public) artwork. They have a high degree of control over the creative process - the choices that go into deciding how to execute the piece - but when it comes to the actual execution itself, they are highly dependent on the community members to carry out the vision. 

\subsection{A Patron of the Arts}

This is someone who has a grand vision for a piece of artwork and commissions an artist to execute it. The final artwork is intended to embody this person’s ideals, and they spend a considerable amount thinking about it, and many in art circles recognize them for their patronage. While they have a high degree of intentionality, they must cede some control to the artist, and they expend a moderate amount of effort. The artist is entirely responsible for the production, and the patron is considerably dependent on the artist for executing the vision, and must cede some of the conceptualization to the artist. 

\section{Creative Ownership Framework Applications}\label{sec:applications}

\begin{figure}[H]
    \centering 
    \includegraphics[width=\textwidth]{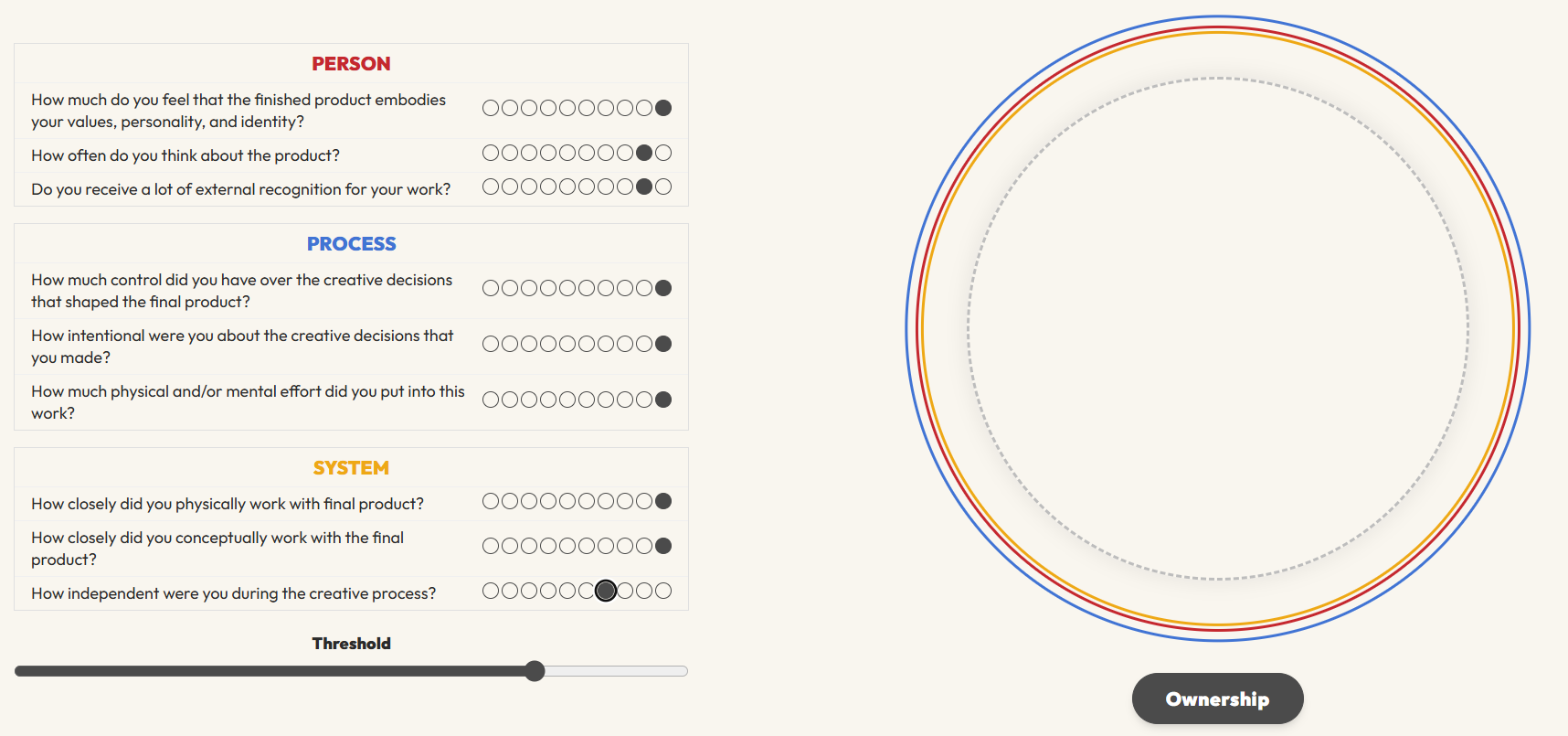} 
    \caption{Ownership Framework for a Master Artist}
    \label{fig:appendix-image1}
\end{figure}

\begin{figure}[H]
    \centering 
    \includegraphics[width=\textwidth]{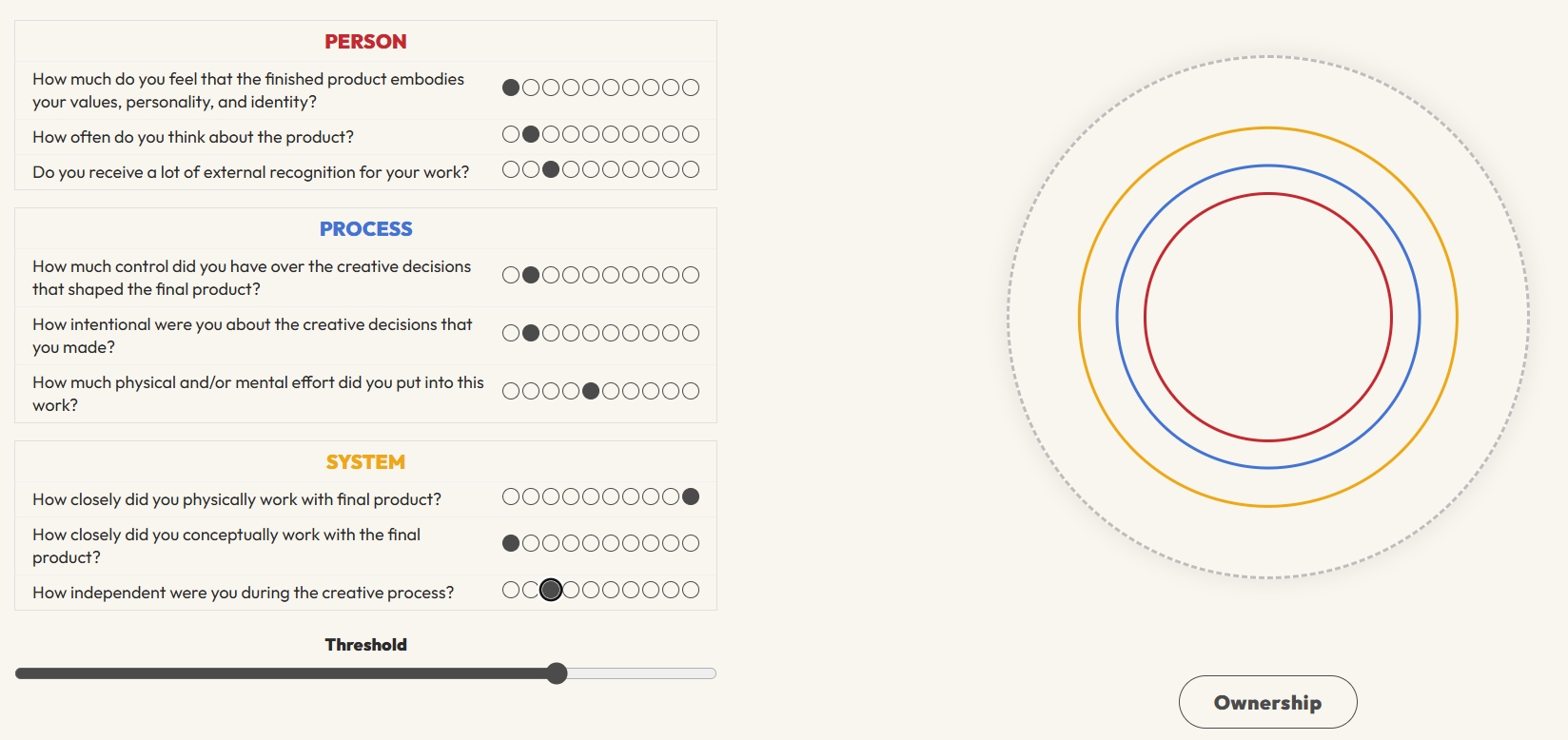}  
    \caption{Ownership Framework for an Amateur Artist}
    \label{fig:appendix-image2}
\end{figure}

\section{Interview Guide}
\label{app:interview}

This appendix reproduces the interview protocol approved by our Institutional Review Board (IRB). Participants were recruited through professional contacts, publicly available sources, and snowball sampling, and were screened to ensure that they: (1) work or participate significantly in a creative field, (2) have at least two finished creative products they are willing to discuss, (3) are fluent in English, and (4) are over 18 years of age.

Interviews lasted approximately 45–60 minutes. At the beginning of each interview, the researcher reiterated the study goals, confirmed that the participant had read the consent form, and obtained verbal consent to record the session. Interviews followed a semi-structured protocol with the following guiding questions and activities:

\begin{enumerate}
    \item Can you tell us about your creative background, current field of work, and expertise?
    \item Can you tell us about two creative products—one associated with low feelings of ownership and one associated with high feelings of ownership—that you have worked on?
    \item Please rank your overall feelings of ownership over each of these projects on a scale of 1–10.
    \item How did you arrive at these ratings? What aspects of the project evoke these feelings?
    \item At this step, participants will be asked to fill out the Creative Ownership Test survey for each of the creative products they have worked on.
    \item Do you feel that the results of this test align with your original ownership feelings?
    \item Are there any dimensions of this framework that feel particularly relevant to your feelings of ownership?
    \item Are there any dimensions that are uninformative, redundant, or not useful toward your feelings of ownership?
    \item Are there any elements you feel are missing from this framework that could better explain your ownership feelings?
\end{enumerate}

 Interviews followed a semi-structured format in which participants answered guiding questions and, where relevant, engaged in follow-up discussion to elaborate on their responses. Data were analyzed anonymously, with only a participant number and their creative practice used as identifiers.

\end{document}